\documentclass{article}
\usepackage[dvips]{epsfig}
\usepackage{LaThuileFPSpro}

\newcommand{\phione}{$\phi^{}_1$}
\newcommand{\phitwo}{$\phi^{}_2$}
\newcommand{\phithree}{$\phi^{}_3$}
\newcommand{\belle}{Belle}

\newcommand{\cp}{$CP$}
\newcommand{\mbc}{$M^{}_{\rm bc}$}
\newcommand{\deltaE}{$\Delta E$}

\newcommand{\apipi}{${\cal A}^{}_{\pi\pi}$}
\newcommand{\spipi}{${\cal S}^{}_{\pi\pi}$}

\newcommand{\bbar}{\overline{B}{}^{\,0}}
\newcommand{\dbar}{\overline{D}{}^{\,0}}
\newcommand{\kbar}{\overline{K}{}^{\,0}}
\newcommand{\bbbar}{$B^0$-$\bbar$}

\newcommand{\ra}{\!\rightarrow\!}
\newcommand{\bjpsiks}{$B^0\ra J/\psi K^0_S$}
\newcommand{\bpipi}{$B^0\ra\pi^+\pi^-$}

\def\simge{\mathrel{%
   \rlap{\raise 0.511ex \hbox{$>$}}{\lower 0.511ex \hbox{$\sim$}}}}
\def\simle{\mathrel{
   \rlap{\raise 0.511ex \hbox{$<$}}{\lower 0.511ex \hbox{$\sim$}}}}

\def\plusominus{\mathrel{%
   \rlap{\raise 0.651ex \hbox{\footnotesize $+$}}
   {\lower 0.651ex \hbox{\footnotesize $-$}}}}
\def\minusoplus{\mathrel{%
   \rlap{\raise 0.651ex \hbox{\footnotesize $-$}}
   {\lower 0.651ex \hbox{\footnotesize $+$}}}}
\begin{document}

\title{ 
\flushright{\rm UCHEP-05-04}
\vskip0.20in
\begin{center}
ANGLES OF THE CKM UNITARITY TRIANGLE \\
MEASURED AT BELLE
\end{center}
  }
\author{
  A.\ J.\ Schwartz    \\
  {\em  Physics Department, University of Cincinnati} \\
  {\em  P.O.\ Box 210011, Cincinnati, Ohio 45221 USA}  \\
  }
\maketitle

\baselineskip=11.6pt

\begin{abstract}
The \belle\ experiment has used several methods to measure or 
constrain the angles $\phi^{}_1,\,\phi^{}_2$, and $\phi^{}_3$
(or $\beta,\,\alpha$, and $\gamma$) 
of the CKM unitarity triangle. The results are
$\sin2\phi^{}_1=0.728\,\pm0.056\,({\rm stat})\,\pm 0.023\,({\rm syst})$
or $\phi^{}_1=\left(23.4\,^{+2.7}_{-2.4}\right)^\circ$ 
from $B^0\ra J/\psi\,K^0$ decays (140~fb$^{-1}$);
$\phi^{}_2=(0\!-\!19)^\circ$ or $(71\!-\!180)^\circ$ 
at 95.4\% CL from \bpipi\ decays (253~fb$^{-1}$); and
$\phi^{}_3=\left[\,68\,^{+14}_{-15}\,({\rm stat})\,\pm 13\,({\rm syst})
\,\pm 11\,({\rm model})\,\right]^\circ$ 
from $B^\pm\ra (D^0,\dbar) K^\pm,\,(D^0,\dbar)\ra K^0_S\,\pi^+\pi^-$ 
decays (253~fb$^{-1}$). These values satisfy the triangle relation 
$\phi^{}_1+\phi^{}_2+\phi^{}_3=180^\circ$ within their uncertainties.
The angle \phione\ is also determined from several
$b\ra s\bar{q}q$ penguin-dominated decay modes; the value
obtained by taking a weighted average of the individual
results differs from the $B^0\ra J/\psi\,K^0$ result 
by more than two standard deviations. The angle \phitwo\ 
is constrained by measuring a \cp\ asymmetry in the decay 
time distribution; the asymmetry observed is large, and 
the difference in the yields of $B^0,\bbar\ra\pi^+\pi^-$ 
decays constitutes the first evidence for direct \cp\ 
violation in the $B$ system.
\end{abstract}

\newpage


\section{Introduction}

The Standard Model predicts \cp\ violation to occur in $B^0$
meson decays owing to a complex phase in the $3\!\times\!3$
Cabibbo-Kobayashi-Maskawa (CKM) mixing matrix\cite{KM}. This 
phase is illustrated by plotting the unitarity condition
$V^*_{ub} V^{}_{ud} + V^*_{cb} V^{}_{cd} + V^*_{tb} V^{}_{td} =0$ 
as vectors in the complex plane: the phase results in 
a triangle of nonzero height.  
Various measurements in the $B$ system are sensitive
to the internal angles \phione, \phitwo, and \phithree\ 
(also known as $\beta$, $\alpha$, and $\gamma$,
respectively); these measurements allow us to determine 
the angles and check whether the triangle closes. 
Non-closure would indicate physics beyond the Standard Model.
Here we present measurements of \phione\ and \phitwo\ obtained by 
measuring time-dependent \cp\ asymmetries, and a measurement of
\phithree\ obtained by measuring an asymmetry in the Dalitz plot 
distribution of three-body decays. The results presented are from 
the Belle experiment\cite{belle}, which runs at the KEKB 
asymmetric-energy $e^+e^-$ collider\cite{kekb}
operating at the $\Upsilon(4S)$ resonance.

In Belle, pion and kaon tracks are identified using 
information from time-of-flight counters, aerogel 
\v{C}erenkov counters, and $dE/dx$ information from 
the central tracker\cite{bellePID}. $B$ decays are 
identified using the ``beam-constrained'' mass
$M^{}_{\rm bc}\equiv\sqrt{E^2_{\rm beam}-p^2_B}$ and 
the energy difference $\Delta E\equiv E^{}_B-E^{}_{\rm beam}$,
where $p^{}_B$ is the reconstructed $B$ momentum, 
$E^{}_B$ is the reconstructed $B$ energy, and $E^{}_{\rm beam}$ is 
the beam energy, all evaluated in the $e^+e^-$ center-of-mass (CM) frame. 
A tagging  algorithm\cite{tagging} is used to identify the flavor at 
production of the decaying~$B$, i.e., whether it is $B^0$ or~$\bbar$.
This algorithm examines tracks not associated with the 
signal decay to identify the flavor of the non-signal~$B$. 
The signal-side tracks are fit for a decay vertex, and the 
tag-side tracks are fit for a separate decay vertex; the 
distance $\Delta z$ between vertices is to a very good approximation 
proportional to the time difference $\Delta t$ between the $B$ 
decays: $\Delta z\approx (\beta\gamma c)\Delta t$, where
$\beta\gamma$ is the Lorentz boost of the CM system.

The dominant background is typically
$e^+e^-\!\ra q\bar{q}$ continuum events, where
$q=u,d,s,c$. In the CM frame such events tend 
to be jet-like, whereas $B\overline{B}$ events 
tend to be spherical. The sphericity of an event
is usually quantified via
Fox-Wolfram moments\cite{fox_wolfram} of the form
$h^{}_\ell = \sum_{i,j} p^{}_i\,p^{}_j\,P^{}_\ell(\cos\theta^{}_{ij})$,
where $i$ runs over all tracks on the tagging side and
$j$ runs over all tracks on either the tagging side or
the signal side\cite{KSFW}. The function $P^{}_\ell$ is 
the $\ell$th Legendre polynomial and $\theta^{}_{ij}$ is 
the angle between momenta $\vec{p^{}_i}$ and $\vec{p^{}_j}$ 
in the CM frame. These moments are combined into a Fisher 
discriminant, and this is combined with the probability
density function (PDF) for $\cos\theta^{}_B$, where $\theta^{}_B$ 
is the polar angle in the CM frame between the $B$ direction and 
the $z$ axis (nearly along the $e^-$ beam direction).
$B\overline{B}$ events are produced 
with a $1-\cos^2\theta^{}_B$ distribution while $q\bar{q}$ events 
are produced uniformly in $\cos\theta^{}_B$. The PDFs for signal 
and $q\bar{q}$ background are obtained using MC simulation and 
\mbc-\deltaE\ sidebands in data, respectively. We use the products 
of the PDFs to calculate a signal likelihood ${\cal L}^{}_s$ and 
a continuum likelihood ${\cal L}^{}_{q\bar{q}}$ and require that 
${\cal L}^{}_s/({\cal L}^{}_s + {\cal L}^{}_{q\bar{q}})$
be above a threshold. 


The angles \phione\ and \phitwo\ are determined by measuring the 
time dependence of decays to \cp-eigenstates. This distribution 
is given by
\begin{eqnarray}
\hspace*{-0.12in} \frac{dN}{d\Delta t}\hspace*{-0.08in} & 
\propto\  & \hspace*{-0.12in} 
e^{-\Delta t/\tau}\biggl[1 -q\Delta\omega + q(1-2\omega)
\left[\,{\cal A} \cos(\Delta m\Delta t) +
 {\cal S} \sin(\Delta m \Delta t)\,\right]\,\biggr], 
\label{eqn:master1}
\end{eqnarray}
where $q\!=\!+1$ ($-1$) corresponds to $B^0$ ($\bbar$) 
tags, $\omega$ is the mistag probability, 
$\Delta\omega$ is a possible difference in $\omega$
between $B^0$ and $\bbar$ tags, and $\Delta m$ 
is the \bbbar\ mass difference. The \cp-violating coefficients 
${\cal A}$ and ${\cal S}$ are functions of the parameter $\lambda$:
${\cal A}=(|\lambda|^2-1)/(|\lambda|^2+1)$ and
${\cal S}=2{\rm\,Im}(\lambda)/(|\lambda|^2+1)$, where
\begin{eqnarray}
\hskip-0.10in \lambda & \hskip-0.05in = & \hskip-0.05in
\frac{q}{p}\,\frac{A(\bbar\ra f)}{A(B^0\ra f)}\ \approx\ 
\sqrt{\frac{M^*_{12}}{M^{}_{12}}}\,\frac{A(\bbar\ra f)}{A(B^0\ra f)}\ =\ 
\left(\frac{V^{}_{td}\,V^*_{tb}}{V^*_{td}\,V^{}_{tb}}\right)\,
\frac{A(\bbar\ra f)}{A(B^0\ra f)}\,.
\label{eqn:master2}
\end{eqnarray}
In this expression, $q$ and $p$ are the complex coefficients relating 
the flavor eigenstates $B^0$ and $\bbar$ to the mass eigenstates, 
$M^{}_{12}$ is the off-diagonal element of the $B^0$-$\bbar$ mass 
matrix, and we assume that the off-diagonal element of the decay 
matrix is much smaller: $\Gamma^{}_{12}\ll M^{}_{12}$. If only one 
weak phase enters the decay amplitude $A(\bbar\ra f)$, then 
$|A(\bbar\ra f)/A(B^0\ra f)|=1$ and $\lambda=\eta^{}_f\,e^{i\,2\theta}$,
where $\eta^{}_f\!=\!\pm 1$ is the \cp\ of the final state~$f$.
For the final states discussed here, $|\theta|=\phi^{}_1$ or~$\phi^{}_2$.

\section{The angle {\boldmath $\phi^{}_1$}}

This angle is most accurately measured using $B^0\ra J/\psi\,K^0$ 
decays\footnote{This measurement includes $B^0\ra J/\psi\,K^0_S$,
$J/\psi\,K^0_L$, $\psi(2S)K^0_S$, $\chi^{}_{c1}K^0_S$, $\eta^{}_c K^0_S$, 
and $J/\psi\,K^{*\,0}\,(K^{*\,0}\ra K^0_S\pi^0)$; we use 
``$B^0\ra J/\psi\,K^0\,$'' to denote all six modes.}.
The decay is dominated by a $b\ra c\bar{c}s$ tree amplitude 
and a $b\ra s\bar{c}c$ penguin amplitude. The latter can be 
divided into two pieces: a piece with $c$ and $t$ in the
loop that has the same weak phase as the tree amplitude, 
and a piece with $u$ and $t$ in the loop that has a different
weak phase but is suppressed by $\sin^2\theta^{}_C$ relative 
to the first piece. Due to this suppression, $A(\bbar\ra f)$ 
is governed by a single weak phase: ${\rm Arg}(V^{}_{cb}\,V^*_{cs})$.
The ratio $A(\bbar\ra J/\psi\,K^0_S)/A(B^0\ra J/\psi\,K^0_S)$
includes an extra factor 
$(p/q)^{}_K= V^*_{cd}\,V^{}_{cs}/(V^{}_{cd}\,V^*_{cs})$ 
to account for the $\kbar$ oscillating to a $K^0_S$, and thus
$\lambda = -\left[V^{}_{td}\,V^*_{tb}/(V^*_{td}\,V^{}_{tb})\right]
\left[V^{}_{cb}\,V^*_{cs}/(V^*_{cb}\,V^{}_{cs})\right]
\left[V^*_{cd}\,V^{}_{cs}/(V^{}_{cd}\,V^*_{cs})\right]
 = -e^{-i\,2\phi^{}_1}$. The \cp\ asymmetry parameters are 
therefore ${\cal S}=\sin2\phi^{}_1$, ${\cal A}=0$. To determine 
\phione, we fit the $\Delta t$ distribution for ${\cal S}$; 
the result is
$\sin2\phi^{}_1=0.728\,\pm0.056\,({\rm stat})\,\pm 0.023\,({\rm syst})$,
or $\phi^{}_1=\left(23.4\,^{+2.7}_{-2.4}\right)^\circ$
(the smaller of the two solutions for $\phi^{}_1$). 
The fit result for ${\cal A}$ yields
$|\lambda| = 1.007\,\pm 0.041\,({\rm stat})\,\pm 0.033\,({\rm syst})$,
in agreement with the theoretical expectation. These results correspond 
to 140~fb$^{-1}$ of data\cite{belle_sin2phi1}.

\begin{figure}[t]
\begin{center}
  \mbox{\epsfig{file=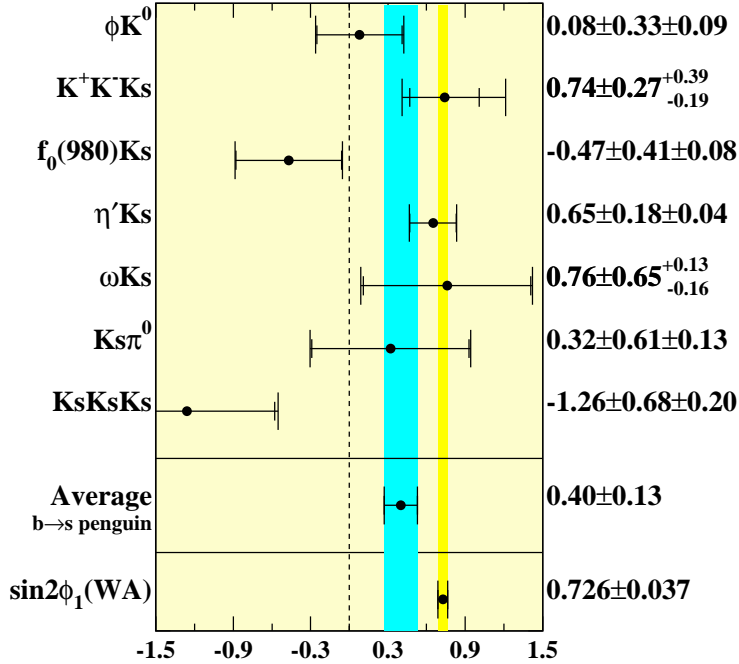,width=3.8in}}
\end{center}
\vspace{-0.10in}
\caption{\it Values of $\sin2\phi^{}_1$ measured
in decay modes dominated by $b\ra s\bar{q}q$ penguin 
amplitudes, for 253~fb$^{-1}$ of data. The average 
value differs from the world average (WA) value 
measured in $B^0\ra J/\psi\,K^0$ decays. }
\label{fig:beta}
\end{figure}

There are several decay modes that proceed exclusively via
penguin amplitudes (e.g., $\bbar\ra\phi\,\kbar$ proceeding via
$b\ra s\bar{s}s$) or else are dominated by penguin amplitudes
(e.g., $\bbar\ra (\eta'/\omega/\pi^0)\kbar$ proceeding via 
$b\ra s\bar{d}d$) but have the same weak phase as the
$b\ra c\bar{c}s$ tree amplitude. 
This is because the penguin loop factorizes 
into a $c,t$ loop with the same weak phase 
and a $u,t$ loop with a different weak phase;
the latter, however, is suppressed by $\sin^2\theta^{}_C$ 
relative to the former and plays a negligible role.
We thus expect these decays to also have 
${\cal S}=\sin2\phi^{}_1$, ${\cal A}= 0$. 
There are small mode-dependent corrections 
($|\Delta{\cal S}|\leq 0.10$) to this prediction 
due to final-state rescattering\cite{soni_fsi}. 
Table~\ref{tab:beta} lists these 
modes and the corresponding values\cite{belle_bsqq} of 
$\sin2\phi^{}_1$ obtained from fitting the $\Delta t$ 
distributions; Fig.~\ref{fig:beta} shows these results 
in graphical form. Neglecting the small rescattering 
corrections and simply averaging the penguin-dominated 
values gives $\sin2\phi^{}_1=0.40\,\pm0.13$. This value
differs from the $B^0\ra J/\psi\,K^0$ world average value 
by 2.4 standard deviations, which may be a statistical 
fluctuation or may indicate new physics.

\begin{table}[t]
\centering
\caption{\it Decay modes used to measure $\sin2\phi^{}_1$, the
number of candidate events, the value of $\sin2\phi^{}_1$ obtained,
and the parameter ${\cal A}$ obtained [see Eq.~(\ref{eqn:master1})].
The $B^0\ra J/\psi\,K^0$ result corresponds to 140~fb$^{-1}$ of data;
the other results correspond to 253~fb$^{-1}$ of data.}
\vskip 0.1 in
\renewcommand{\arraystretch}{1.14}
\begin{tabular}{|lccc|}
\hline
(\cp)\ Mode & 
Candidates & $\sin2\phi^{}_1$ & ${\cal A}$\\
\hline
\hline\hskip-0.10in
\begin{tabular}{l}
$(-)\,J/\psi\,K^0_S$ \\
$(+)\,J/\psi\,K^0_L$
\end{tabular}            & 
\begin{tabular}{c}
2285 \\
2332
\end{tabular} &  $0.728\,\pm\,0.056\,\pm\,0.023$ & -- \\
\hline\hskip-0.10in
\begin{tabular}{l}
$(-)\,\phi\,K^0_S$ \\
$(+)\,\phi\,K^0_L$
\end{tabular}            & 
\begin{tabular}{c}
$139\,\pm 14$ \\
$36\,\pm 15$
\end{tabular} &  $0.08\,\pm\,0.33\,\pm\,0.09$ & $0.08\,\pm 0.22\,\pm 0.09$ \\
\hline
$(\pm)\,K^+K^-K^0_S$ & $398\,\pm 28$ & 
$0.74\,\pm\,0.27\,^{+0.39}_{-0.19}$ & $-0.09\,\pm 0.12\,\pm 0.07$ \\
\ \ ($+=\,83\%$) & & & \\
$(+)\,f^{}_0(980)\,K^0_S$ & $94\,\pm 14$ & 
$-0.47\,\pm\,0.41\,\pm\,0.08$ & $-0.39\,\pm 0.27\,\pm 0.09$ \\
$(+)\,K^0_SK^0_SK^0_S$ & $88\,\pm 13$ & 
$-1.26\,\pm\,0.68\,\pm 0.20$ & $0.54\,\pm 0.34\,\pm 0.09$ \\
\hline
$(-)\,\eta'\,K^0_S$ & $512\,\pm 27$ & 
$0.65\,\pm\,0.18\,\pm\,0.04$ & $-0.19\,\pm 0.11\,\pm 0.05$ \\
$(-)\,\pi^0\,K^0_S$ & $247\,\pm 25$ & 
$0.32\,\pm\,0.61\,\pm 0.13$ & $-0.11\,\pm 0.20\,\pm 0.09$ \\
$(-)\,\omega\,K^0_S$ & $31\,\pm 7$ & 
$0.76\,\pm\,0.65\,^{+0.13}_{-0.16}$ & $0.27\,\pm 0.48\,\pm 0.15$ \\
\hline
\end{tabular}
\label{tab:beta}
\end{table}

\section{The angle {\boldmath $\phi^{}_2$}}

This angle is measured by fitting the $\Delta t$ distribution 
of $B^0\ra\pi^+\pi^-$ decays. The rate is dominated by a 
$b\ra u\bar{u}d$ tree amplitude with a weak phase 
${\rm Arg}(V^{}_{ub}\,V^*_{ud})$. If only this phase were present, 
then $\lambda = \left[V^{}_{td}\,V^*_{tb}/(V^*_{td}\,V^{}_{tb})\right]
\left[V^{}_{ub}\,V^*_{ud}/(V^*_{ub}\,V^{}_{ud})\right] = e^{i\,2\phi^{}_2}$,
and ${\cal S}=\sin2\phi^{}_2$, ${\cal A}=0$. However, a $b\ra d\bar{u}u$ 
penguin amplitude also contributes, and, unlike the penguin in 
\bjpsiks\ decays, the piece with a different weak phase is not
CKM-suppressed relative to the piece with the same weak phase.
The \cp\ asymmetry parameters are therefore more 
complicated\cite{gronau_rosner}:
\begin{eqnarray}
{\cal A}^{}_{\pi\pi} \hskip-0.10in & = & \hskip-0.10in -\,\frac{1}{R}\cdot
\left( 2\left|\frac{P}{T}\right|\sin(\phi^{}_1 + \phi^{}_2)\sin\delta\right)
\label{eqn:cpipi} \\
 & & \nonumber \\
{\cal S}^{}_{\pi\pi} \hskip-0.10in & = & \hskip-0.10in
\frac{1}{R}\cdot
\biggl( 2\left|\frac{P}{T}\right|\sin(\phi^{}_1-\phi^{}_2)\cos\delta +
 \sin 2\phi^{}_2 - \left|\frac{P}{T}\right|^2\sin 2\phi^{}_1 \biggr)
\label{eqn:spipi} \\
 & & \nonumber \\
R \hskip-0.10in & = & \hskip-0.10in 
1-2\left|\frac{P}{T}\right|\cos(\phi^{}_1+\phi^{}_2)\cos\delta + 
\left|\frac{P}{T}\right|^2\,,
\end{eqnarray}
where $|P/T|$ is the magnitude of the penguin amplitude 
relative to that of the tree amplitude, $\delta$ is the 
strong phase difference between the two amplitudes, and
$\phi^{}_1$ is known from $B^0\ra J/\psi K^0$ decays.
Since Eqs.~(\ref{eqn:cpipi}) and (\ref{eqn:spipi}) have three
unknown parameters, measuring \apipi\ and \spipi\ determines 
a volume in $\delta$-$|P/T|$-$\phi^{}_2$ space.

\begin{figure}[t]
\begin{center}
\mbox{\epsfig{file=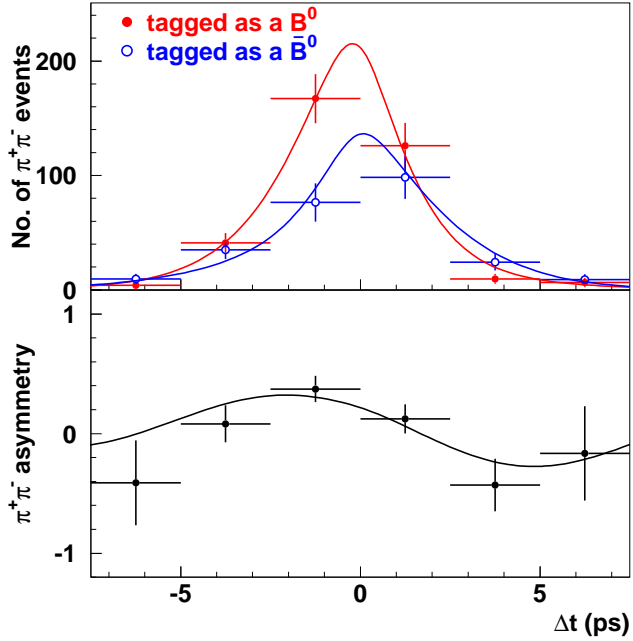,width=3.25in}}
\end{center}
\caption{\it The $\Delta t$ distribution of background-subtracted 
$B^0,\bbar\ra\pi^+\pi^-$ candidates (top), and the resulting \cp\ 
asymmetry $[N(\bbar)\!-\!N(B^0)]/[N(\bbar)\!+\!N(B^0)]$ (bottom).
The smooth curves are projections of the unbinned ML fit. }
\label{fig:pipi} 
\end{figure}

The most recent \belle\ measurement\cite{belle_pipi}
uses 253~fb$^{-1}$ of data; the event sample consists 
of $666\,\pm 43$ $B^0\ra\pi^+\pi^-$ candidates after 
background subtraction. These events are subjected to 
an unbinned maximum likelihood (ML) fit for~$\Delta t$; 
the results are
${\cal A}^{}_{\pi\pi}\!=\!0.56\,\pm\,0.12\,({\rm stat})\,\pm\,0.06\,({\rm syst})$ and
${\cal S}^{}_{\pi\pi}\!=\!-0.67\,\pm\,0.16\,({\rm stat})\,\pm\,0.06\,({\rm syst})$,
which together indicate large \cp\ violation. The nonzero value for
${\cal A}^{}_{\pi\pi}$ indicates {\it direct\/} \cp\ violation.
Fig.~\ref{fig:pipi} shows the $\Delta t$ distributions 
for $q=\pm 1$ tagged events along with projections of the 
ML fit; a clear difference is seen between the fit results.

\begin{figure}[t]
\begin{center}
\mbox{\epsfig{file=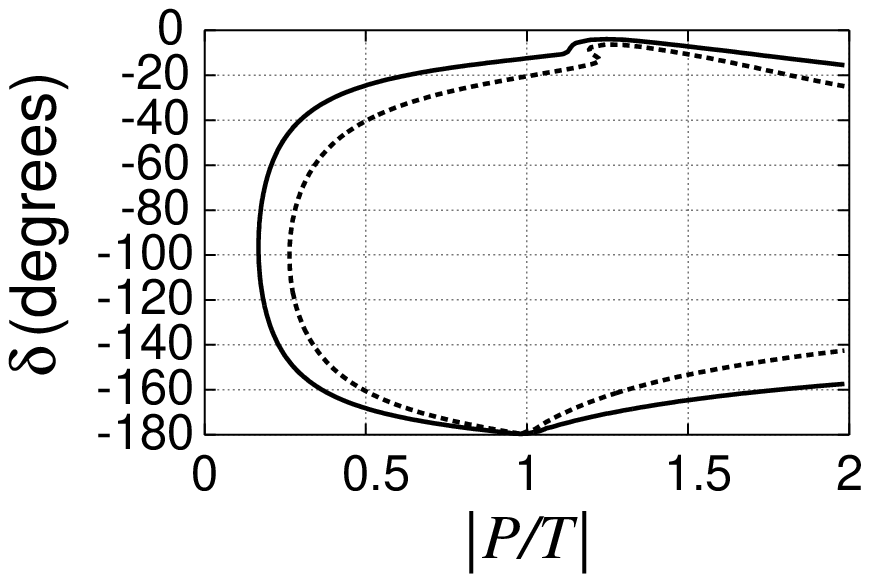,width=4.0in}}
\end{center}
\vskip-0.15in
\caption{\it Projection of the 68.3\%~CL (dashed) and
95.4\%~CL (solid) volumes in $\delta$-$|P/T|$-$\phi^{}_2$ 
space onto the $\delta$-$|P/T|$ axes. From the solid
contour we obtain the constraints $|P/T|\!>\!0.17$ and 
$-180^\circ\!<\!\delta\!<\!-4^\circ$ (95.4\%~CL).}
\label{fig:delta_vs_povert} 
\end{figure}

The values of \apipi\ and \spipi\ determine a 95.4\%~CL ($2\sigma$) 
volume in $\delta$-$|P/T|$-$\phi^{}_2$ space. Projecting this volume 
onto the $\delta$-$|P/T|$ axes gives the region shown in 
Fig.~\ref{fig:delta_vs_povert}; from this region we obtain 
the constraints $|P/T|\!>\!0.17$ for any value of~$\delta$,
and $-180^\circ\!<\!\delta\!<\!-4^\circ$ for any value of~$|P/T|$.

\begin{figure}[t]
\begin{center}
\mbox{\epsfig{file=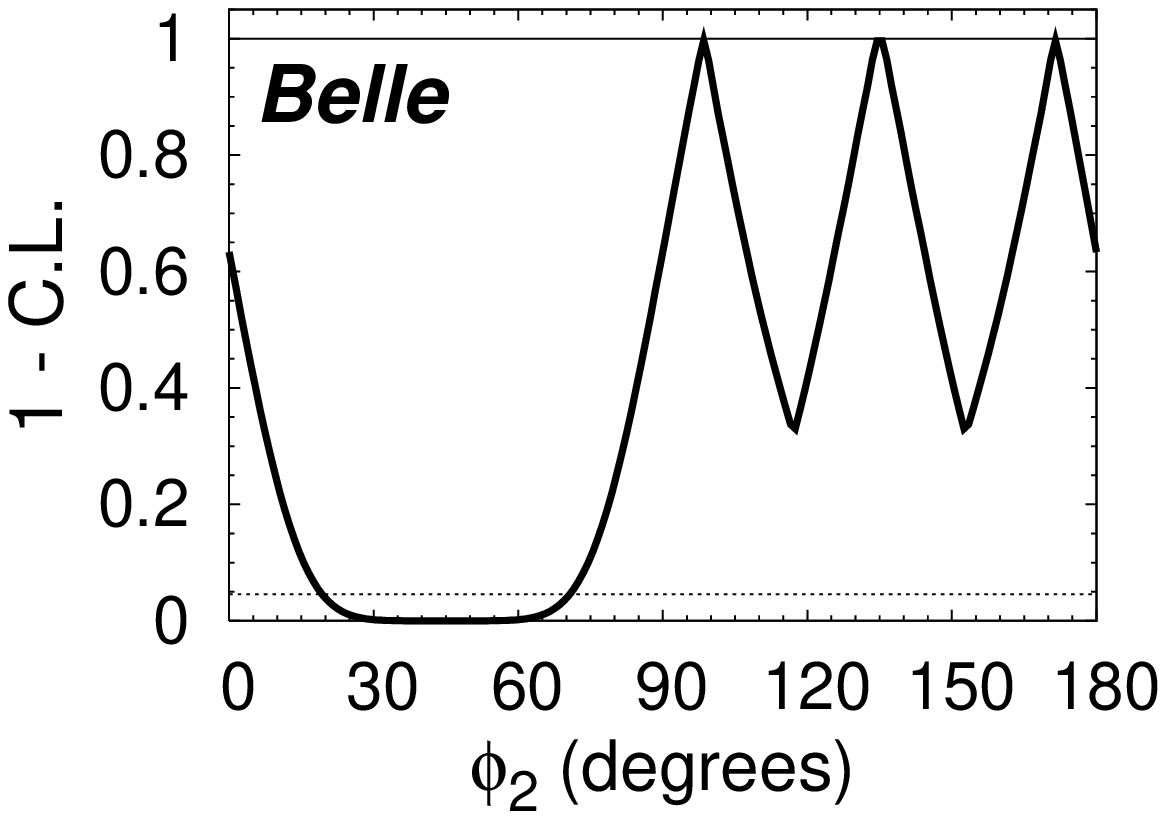,width=3.75in}}
\end{center}
\vskip-0.10in
\caption{\it The result of fitting the branching fractions
for $B\ra\pi^+\pi^-$, $\pi^\pm\pi^0$, $\pi^0\pi^0$ and
the \cp\ asymmetry parameters ${\cal A}^{}_{\pi^+\pi^-}$,
${\cal S}^{}_{\pi^+\pi^-}$, and ${\cal A}^{}_{\pi^0\pi^0}$, 
as a function of~\phitwo\ (see text). The vertical axis
is one minus the confidence level. The horizontal 
line at $1\!-\!{\rm CL}=0.046$ corresponds to a 
95.4\% CL interval for \phitwo.} 
\label{fig:phitwo} 
\end{figure}

The dependence upon $\delta$ and $|P/T|$ can be removed 
by performing an isospin analysis\cite{gronau_london} of 
$B\ra\pi\pi$ decays.
This method uses the measured branching fractions 
for $B\ra\pi^+\pi^-,\,\pi^\pm\pi^0,\,\pi^0\pi^0$ and
the \cp\ asymmetry parameters ${\cal A}^{}_{\pi^+\pi^-}$,
${\cal S}^{}_{\pi^+\pi^-}$, and ${\cal A}^{}_{\pi^0\pi^0}$. 
We scan values of \phitwo\ from $0^\circ\!-\!180^\circ$
and for each value construct a $\chi^2$ based on the 
difference between the predicted values for the six
observables and the measured values. We convert this 
$\chi^2$ into a confidence level (CL) by subtracting
off the minimum $\chi^2$ value and inserting the result
into the cumulative distribution function for the $\chi^2$ 
distribution for one degree of freedom. The resulting 
function $1\!-\!{\rm CL}$ is plotted in Fig.~\ref{fig:phitwo}. 
From this plot we read off a 95.4\%~CL interval 
$\phi^{}_2=(0\!-\!19)^\circ$ or $(71\!-\!180)^\circ$, 
i.e., we exclude the range $20^\circ\!-\!70^\circ$.

\section{The angle {\boldmath $\phi^{}_3$}}

The angle \phithree\ is challenging to measure 
by fitting the $\Delta t$ distribution, 
as the two requisite interfering amplitudes have very 
different magnitudes, and the small ratio of magnitudes 
multiplies the $\phi^{}_3$-dependent term 
$[\sin(2\phi^{}_1\!+\!\phi^{}_3\!\pm\delta)]$\cite{dstarpi}.
As an alternative, one can probe \phithree\ via interference in
the Dalitz plot distribution of $B^\pm\ra (D^0,\dbar)K^\pm$ decays:
the additional phase \phithree\ causes a difference between
the interference pattern for $B^+$ decays and that for 
$B^-$ decays\cite{Bondar}.

We study this asymmetry by reconstructing $B^\pm\ra(D^0,\dbar)K^\pm$ 
decays in which the $D^0$ or $\dbar$ decays to 
the common final state $K^0_S\,\pi^+\pi^-$. 
Denoting $m(K^0_S,\pi^+)\equiv m_+$, $m(K^0_S,\pi^-)\equiv m_-$, 
$A(D^0\ra K^0_S\,\pi^+\pi^-)\equiv A(m^+,m^-)$, and
$A(\dbar\ra K^0_S\,\pi^+\pi^-)\equiv \overline{A}(m^+,m^-)=A(m^-,m^+)$
(i.e., assuming \cp\ conservation in $D^0$ decays), we have
\begin{eqnarray}
A(B^+\ra\tilde{D}^0 K^+,\,\tilde{D}^0\ra K^0_S\,\pi^+\pi^-) & = &
    A(m^2_+,m^2_-) + r e^{i(\delta+\phi^{}_3)}A(m^2_-,m^2_+) \nonumber \\
 & & \\
A(B^-\ra\tilde{D}^0 K^-,\,\tilde{D}^0\ra K^0_S\,\pi^+\pi^-) & = &
    A(m^2_-,m^2_+) + r e^{i(\delta-\phi^{}_3)}A(m^2_+,m^2_1)\,, \nonumber \\
 & & 
\end{eqnarray}
where $\tilde{D}^0$ denotes $(D^0+\dbar)$, $r$ is 
the ratio of magnitudes of the two amplitudes 
$\left|A(B^+\ra D^0 K^+)/A(B^+\ra\dbar K^+)\right|$, 
and $\delta$ is the strong phase difference between 
the amplitudes. The decay rates are given by
\begin{eqnarray}
\left|A(B^+\ra\tilde{D}^0 K^+\ra (K^0_S\,\pi^+\pi^-) K^+)\right|^2\! 
&\!=\! &\! |A(m^2_+,m^2_-)|^2 + r^2 |A(m^2_-,m^2_+)|^2 + \nonumber \\
 & & \hskip-0.90in 
2r|A(m^2_+,m^2_-)||A(m^2_-,m^2_+)|\cos(\delta+\phi^{}_3+\theta) 
\label{eqn:dalitz1} \\
 & & \nonumber \\
\left|A(B^-\ra\tilde{D}^0 K^-\ra (K^0_S\,\pi^+\pi^-) K^-)\right|^2 \!
& \!=\! &\! r^2|A(m^2_+,m^2_-)|^2 + |A(m^2_-,m^2_+)|^2 +\nonumber \\
 & & \hskip-0.90in 
2r|A(m^2_+,m^2_-)||A(m^2_-,m^2_+)|\cos(\delta-\phi^{}_3+\theta)\,,
\label{eqn:dalitz2} 
\end{eqnarray}
where $\theta$ is the phase difference between $A(m^2_+\,,m^2_-)$
and $A(m^2_-\,,m^2_+)$ and varies over the Dalitz plot. 
Thus, given a $D^0\ra K^0_S\,\pi^+\pi^-$ decay model 
$A(m^2_+,m^2_-)$, one can fit
the $B^\pm$ Dalitz plots to Eqs.~(\ref{eqn:dalitz1}) 
and (\ref{eqn:dalitz2}) to determine the parameters 
$r,\,\delta$, and~\phithree. The decay model is 
determined from data, i.e., $D^0\ra K^0_S\,\pi^+\pi^-$ 
decays produced via $e^+e^-\!\ra c\bar{c}$.

\begin{figure}[tb]
\hbox{\hskip-0.10in\vbox{
\mbox{\epsfig{file=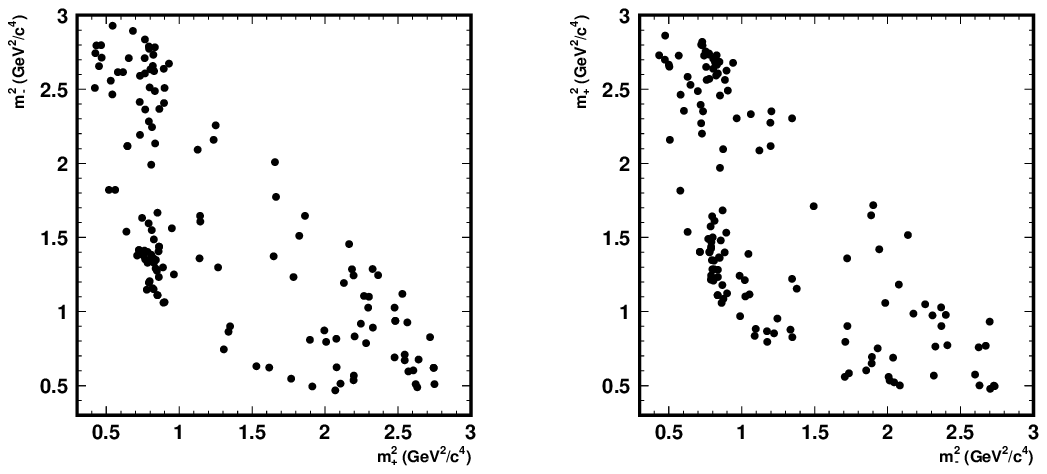,width=4.9in}}
\vskip0.30in
\mbox{\epsfig{file=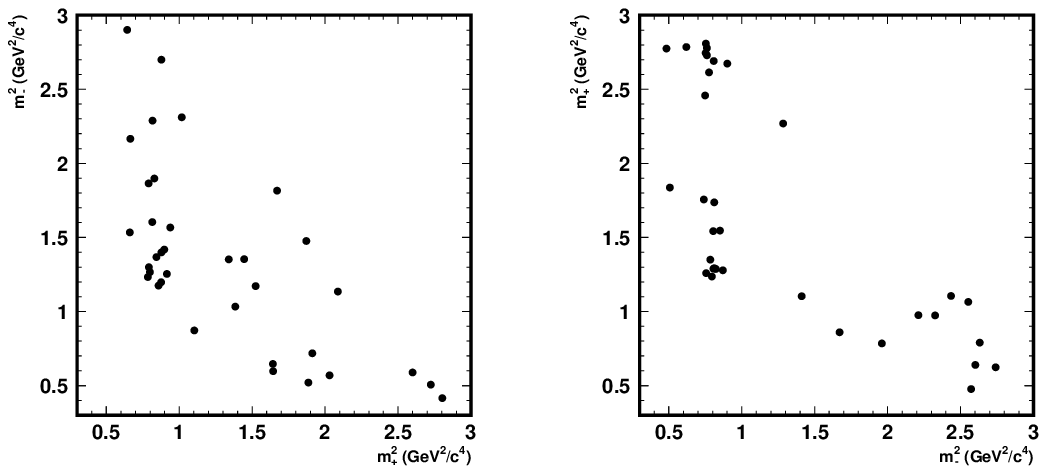,width=4.9in}}
}}
\vskip0.20in
\caption{\it Dalitz plots of $\tilde{D}^0\ra K^0_S\,\pi^+\pi^-$
decays obtained from samples of 
$B^+\ra\tilde{D}^0 K^+$ (top left), 
$B^-\ra\tilde{D}^0 K^-$ (top right), 
$B^+\ra\tilde{D}^{*\,0} K^+$ (bottom left), and
$B^-\ra\tilde{D}^{*\,0} K^-$ (bottom right).}
\label{fig:dalitzplots} 
\end{figure}

The data sample used consists of 
253~fb$^{-1}$; there are $209\,\pm 16$ $B^\pm\ra\tilde{D}^0 K^\pm$ 
candidates with 75\% purity, and an additional $58\,\pm 8$
$B^\pm\ra\tilde{D}^0{}^* K^\pm\ (\tilde{D}^0{}^*\ra\tilde{D}^0\pi^0)$ 
candidates with 87\% purity\cite{belle-conf-0476}. The background is 
dominated by $q\bar{q}$ continuum events in which a real $D^0$ is 
combined with a random kaon, and random combinations of tracks in 
continuum events. The Dalitz plots for the final samples 
are shown in Fig.~\ref{fig:dalitzplots}. 

The events are subjected to an unbinned ML fit for
$r,\,\delta$, and \phithree. The decay model is a 
coherent sum of two-body amplitudes and a constant
term for the nonresonant contribution:
\begin{eqnarray}
A(m^2_+\,,m^2_-) & = & \sum_{j=1}^N
a^{}_j\,e^{i \alpha^{}_j} {\cal A}^{}_j(m^2_+\,,m^2_-)\ +\ 
a^{}_{\rm nonres}\,e^{i \alpha^{}_{\rm nonres}}\,,
\end{eqnarray}
where $a^{}_j,\,\alpha^{}_j$, and ${\cal A}^{}_j$ are 
the magnitude, phase, and matrix element, respectively, 
of resonance $j$; and $N\!=\!18$ resonances are considered. 
The parameters $a^{}_j$ and $\alpha^{}_j$ are determined by
fitting a large sample of continuum $D^0\ra K^0_S\,\pi^+\pi^-$ 
decays. The dominant intermediate modes\cite{belle-conf-0476}
as determined from the fraction 
$\int |a^{}_j{\cal A}^{}_j|^2\,dm^2_+\,dm^2_- /
\int |A(m^2_+\,,m^2_-)|^2\,dm^2_+\,dm^2_-$
are $K^*(892)^+\pi^-$ (61.2\%), $K^0_S\,\rho^0$ (21.6\%), 
nonresonant $K^0_S\,\pi^+\pi^-$ (9.7\%), and
$K^*_0(1430)^+\pi^-$ (7.4\%).

\begin{figure}[tb]
\hbox{\hskip-0.10in\vbox{
\mbox{\epsfig{file=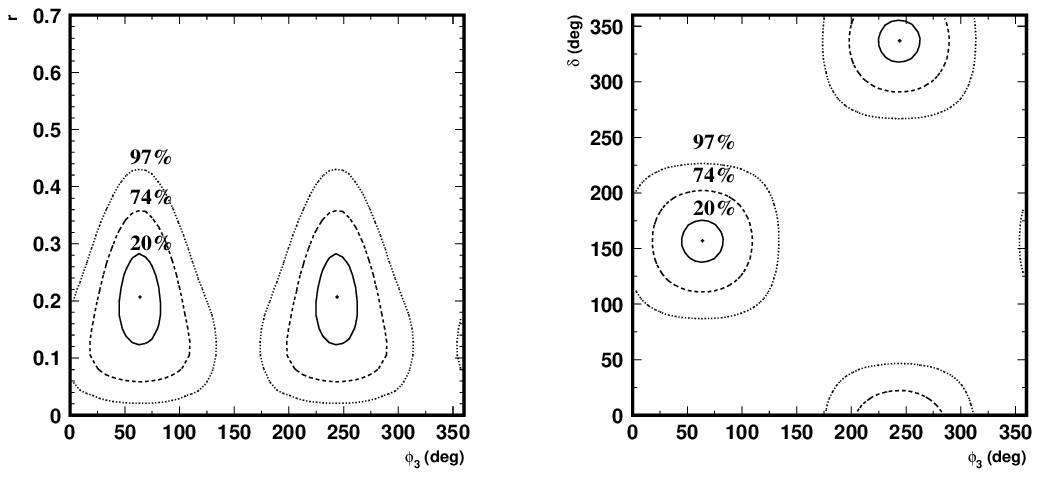,width=4.9in}}
\vskip0.30in
\mbox{\epsfig{file=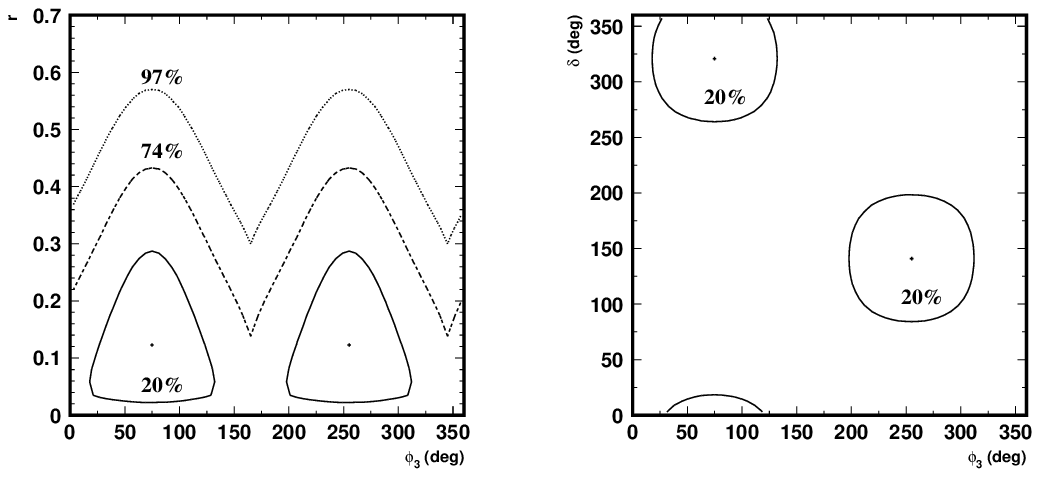,width=4.9in}}
}}
\vskip0.20in
\caption{\it Confidence regions for pairs of parameters:
the left-most plots correspond to $r$-\phithree\ and 
the right-most plots to $\delta$-\phithree. The top 
row corresponds to $B^\pm\ra\tilde{D}^0 K^\pm$ decays
and the bottom row to $B^\pm\ra\tilde{D}^{*\,0} K^\pm$
decays. }
\label{fig:confregions} 
\end{figure}

The central values obtained by the fit are
$r\!=\!0.25,\,\delta\!=\!157^\circ$, and $\phi^{}_3\!=\!64^\circ$ 
for $B^+\ra\tilde{D}^0 K^+$; and $r\!=\!0.25,\,\delta\!=\!321^\circ$, 
and $\phi^{}_3\!=\!75^\circ$ for $B^+\ra\tilde{D}^{*\,0} K^+$. The 
errors obtained by the fit correspond to Gaussian-shaped likelihood
distributions, and for this analysis the distributions are 
non-Gaussian. We therefore use a frequentist MC method to 
evaluate the statistical errors. We first obtain a 
PDF for the fitted parameters $r,\,\delta,\,\phi^{}_3$ 
as a function of the true parameters 
$\bar{r},\,\bar{\delta},\,\bar{\phi}^{}_3$.
We do this by generating several hundred experiments for a given 
set of $\bar{r},\,\bar{\delta},\,\bar{\phi}^{}_3$ values, with each 
experiment having the same number of events as the data, and
fitting these experiments as done for the data. The resulting
distributions for $\alpha^{}_\pm\!=\!r\cos(\delta\pm\phi^{}_3)$ and 
$\beta^{}_\pm\!=\!r\sin(\delta\pm\phi^{}_3)$ are modeled as 
Gaussians ($G$) with mean values $\bar{\alpha}^{}_\pm$ and
$\bar{\beta}^{}_\pm$ and common standard deviation~$\sigma$, 
and the product
$G(\alpha^{}_+\!-\bar{\alpha}^{}_+)\cdot 
 G(\alpha^{}_-\!-\bar{\alpha}^{}_-)\cdot 
 G(\beta^{}_+\!-\bar{\beta}^{}_+)\cdot 
 G(\beta^{}_-\!-\bar{\beta}^{}_-)$ is used to obtain the PDF
${\cal P}(r,\,\delta,\,\phi^{}_3 | \bar{r},\,\bar{\delta},\,\bar{\phi}^{}_3)$.
With this PDF we calculate the confidence level for 
$\{\bar{r},\,\bar{\delta},\,\bar{\phi}^{}_3\}$ given the fit values 
$\{0.25,\,157^\circ,\,64^\circ\}$ for $B^+\ra\tilde{D}^0K^+$ and 
$\{0.25,\,321^\circ,\,75^\circ\}$ for $B^+\ra\tilde{D}^{*\,0}K^+$. 
The resulting confidence regions for pairs of parameters are shown 
in Fig.~\ref{fig:confregions}. The plots show 20\%, 74\%, and 97\%~CL 
regions, which correspond to one, two, and three standard deviations, 
respectively, for a three-dimensional Gaussian distribution. 
The 20\% CL regions are taken as the statistical errors; the
values that maximize the PDF are taken as the central values.
Of the two possible solutions $(\delta,\phi^{}_3)$ or
$(\delta+\pi,\phi^{}_3+\pi)$, we choose the one that
satisfies $0^\circ\!<\!\phi^{}_3\!<\!180^\circ$.

All results are listed in Table~\ref{tab:phi3_results}. 
The second error listed is systematic and results mostly from 
uncertainty in the background Dalitz plot density, variations 
in efficiency, the $m^2_{\pi\pi}$ resolution, and possible 
fitting bias. The third error listed results from uncertainty 
in the $D^0\ra K^0_S\,\pi^+\pi^-$ decay model, e.g., from the 
choice of form factors used for the intermediate resonances
and the $q^2$ dependence of the resonance widths.

\begin{table}[t]
\centering
\caption{\it Results of the 
Dalitz plot analysis for $r,\,\delta$, and \phithree. 
The first error listed is statistical and 
is obtained from a frequentist MC method (see text); the 
second error listed is systematic but does not include uncertainty
from the $\tilde{D}^0\ra K^0_S\,\pi^+\pi^-$ decay model; the 
third error listed is due to the decay model.}
\vskip 0.1 in
\renewcommand{\arraystretch}{1.14}
\begin{tabular}{|c|cc|}
\hline
{\bf Parameter} & {\boldmath $B^+\ra\tilde{D}^0 K^+$}  &  
                      {\boldmath $B^+\ra\tilde{D}^{*\,0} K^+$} \\
\hline
\hline
$r$ & $0.21\,\pm 0.08\,\pm 0.03\,\pm 0.04$ & 
      $0.12\,^{+0.16}_{-0.11}\,\pm 0.02\,\pm 0.04$ \\
$\delta$ & $157^\circ\,\pm 19^\circ\,\pm 11^\circ\,\pm 21^\circ$ & 
           $321^\circ\,\pm 57^\circ\,\pm 11^\circ\,\pm 21^\circ$ \\ 
$\phi^{}_3$ & $64^\circ\,\pm 19^\circ\,\pm 13^\circ\,\pm 11^\circ$ & 
              $75^\circ\,\pm 57^\circ\,\pm 11^\circ\,\pm 11^\circ$ \\
\hline
\end{tabular}
\label{tab:phi3_results}
\end{table}

We combine the $B^+\ra\tilde{D}^0 K^+$ and 
$B^+\ra\tilde{D}^{*\,0} K^+$ results by multiplying together
their respective PDF's, taking the parameter $\bar{\phi}^{}_3$ 
to be common between them. This gives a PDF for the six measured 
parameters $r^{}_1$, $\delta^{}_1$, $\phi^{}_{3\,(1)}$,
$r^{}_2$, $\delta^{}_2$, $\phi^{}_{3\,(2)}$
in terms of the five true parameters
$\bar{r}^{}_1$, $\bar{\delta}^{}_1$,
$\bar{r}^{}_2$, $\bar{\delta}^{}_2$, $\bar{\phi}^{}_3$.
The value of $\bar{\phi}^{}_3$ that maximizes the PDF is 
taken as the central value, and the 3.7\% CL interval prescribed
by the PDF (corresponding to $1\sigma$ for a five-dimensional 
Gaussian distribution) is taken as the statistical error. The 
systematic error is taken from the $B^+\ra\tilde{D}^0 K^+$ 
measurement, as this sample dominates the combined measurement. 
The overall result is
\begin{eqnarray*}
\phi^{}_3 & = & \left[\,68\,^{+14}_{-15}\,({\rm stat})
\,\pm 13\,({\rm syst})\,\pm 11\,({\rm decay\ model})\,\right]^\circ\,.
\end{eqnarray*}
The $2\sigma$ confidence interval including the
systematic error and decay model error is
$22^\circ\!<\!\phi^{}_3\!<\!113^\circ$.

In summary, the Belle experiment has measured or constrained
the angles $\phi^{}_1,\,\phi^{}_2$, and $\phi^{}_3$ of
the CKM unitarity triangle. We obtain
$\sin2\phi^{}_1=0.728\,\pm0.056\,({\rm stat})\,\pm 0.023\,({\rm syst})$
or $\phi^{}_1=\left(23.4\,^{+2.7}_{-2.4}\right)^\circ$ with
140~fb$^{-1}$ of data;
$\phi^{}_2=(0\!-\!19)^\circ$ or $(71\!-\!180)^\circ$ 
at 95.4\% CL with 253~fb$^{-1}$ of data; and
$\phi^{}_3=\left[\,68\,^{+14}_{-15}\,({\rm stat})
\,\pm 13\,({\rm syst})\,\pm 11\,({\rm decay\ model})\,\right]^\circ$ 
with 253~fb$^{-1}$ of data. Within their uncertainties, these values
satisfy the triangle relation $\phi^{}_1+\phi^{}_2+\phi^{}_3=180^\circ$.
The angle \phione\ is measured from $B^0\ra J/\psi\,K^0$ decays and 
also from several $b\ra s\bar{q}q$ penguin-dominated decay modes; 
the value obtained from the penguin modes differs from the 
$B^0\ra J/\psi\,K^0$ result by $2.4\,\sigma$. 
The \phitwo\ constraint results from measuring the \cp\ asymmetry 
coefficients ${\cal A}^{}_{\pi\pi}$ and ${\cal S}^{}_{\pi\pi}$ in 
\bpipi\ decays; the results are
${\cal A}^{}_{\pi\pi}\!=\!0.56\,\pm\,0.12\,({\rm stat})\,\pm\,0.06\,({\rm syst})$ and
${\cal S}^{}_{\pi\pi}\!=\!-0.67\,\pm\,0.16\,({\rm stat})\,\pm\,0.06\,({\rm syst})$, 
which together indicate large \cp\ violation. The nonzero value 
for ${\cal A}^{}_{\pi\pi}$ indicates direct \cp\ violation; the
statistical significance (including systematic uncertainty) 
is~$4.0\,\sigma$. These values also imply 
that the magnitude of the penguin amplitude relative to that of the 
tree amplitude ($|P/T|$) is greater than 0.17 at 95.4\%~CL, and 
that the strong phase difference ($\delta$) lies in the range
$(-180^\circ,-4^\circ)$ at 95.4\%~CL. The \phithree\ measurement 
is obtained from a Dalitz plot analysis of
$B^\pm\ra\tilde{D}^{(*)\,0}K^\pm,\,\tilde{D}^0\ra K^0_S\,\pi^+\pi^-$ 
decays; the statistical significance of the observed (direct) \cp\ 
violation is~98\%.

\section{Acknowledgments}

The author thanks his Belle colleagues for many fruitful
discussions, and the organizers of {\it Les Rencontres de
Physique de la Vall\'{e}e d'Aoste\/} for a well-organized
and stimulating conference.

\end{document}